\documentclass[doublecol]{epl2}
\usepackage{graphicx}
\usepackage[skip=0pt,font=normalsize]{caption}
\usepackage[
  backend=biber,
  style=phys,
  maxbibnames=2,
  articletitle=false
]{biblatex}
\usepackage{amsmath}
\usepackage{mathabx}
\usepackage{commath}
\usepackage{dsfont}
\usepackage{microtype}
\usepackage{flexisym}
\usepackage{hyperref}
\usepackage[export]{adjustbox}
\usepackage{siunitx}

\newcommand{\Dunit}{\si{\micro\metre\squared\per\second}}
\newcommand{\Drunit}{\si{\radian\squared\per\second}}

\definecolor{myGreen}{rgb}{0.1,0.6,0.1}

\title{Feedback control and delayed interactions in active matter}
\shorttitle{Feedback control and delayed interactions in active matter} 

\author{Viktor Holubec \inst{1} \and Frank Cichos \inst{2}}
\shortauthor{V. Holubec and F. Cichos}

\institute{
\inst{1} Charles University, Faculty of Mathematics and Physics, Department of Macromolecular Physics, V Holešovičkách 2, CZ-180 00 Praha \email{viktor.holubec@matfyz.cuni.cz }\\
  \inst{2} Molecular Nanophotonics Group, Peter Debye Institute for Soft Matter Physics, Universit{\"a}t Leipzig
 04103 Leipzig, Germany \email{cichos@physik.uni-leipzig.de}
}

\abstract{
Feedback control plays a central role in active matter, yet it is
inevitably accompanied by noise and finite perception--action delays.
This Perspective reviews recent advances on active systems with delayed
interactions, showing how time delay can induce activity, chirality,
transport, and collective pattern formation, and can act as an effective
control parameter for switching between dynamical states.  We discuss
representative single-particle and many-body systems, highlight key
experimental realizations, and argue that time delay constitutes an
underexplored dimension of morphological intelligence--where intrinsic
response dynamics, rather than explicit sensors or computation, enable
functional behavior in active matter.
}

\begin{document}

\maketitle

\section{Introduction}

\begin{figure}
    \centering
    \includegraphics[width=0.45\textwidth]{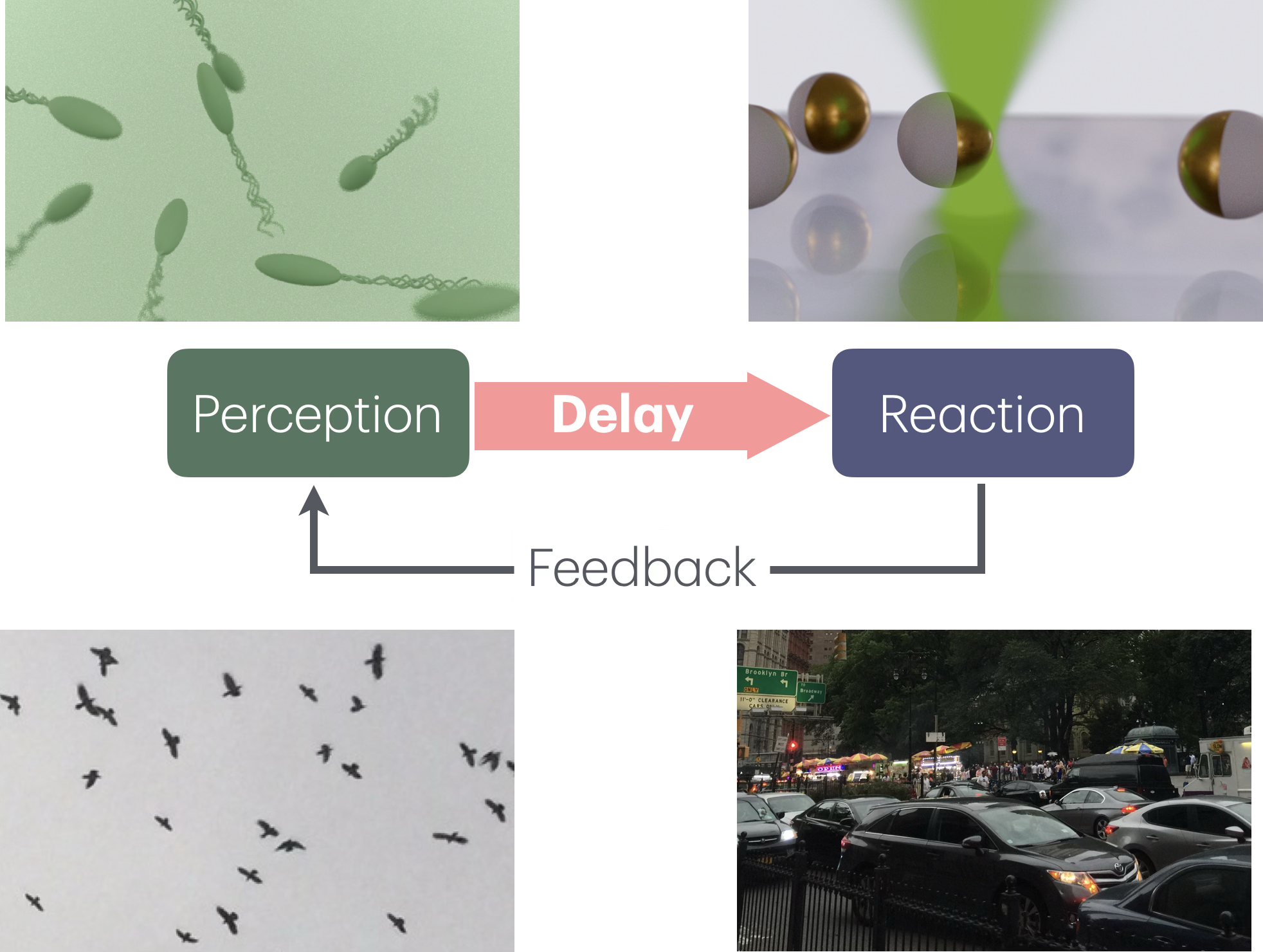}
    \caption{All living systems and most artificial active-matter systems rely on feedback, which inevitably introduces delays into their dynamics. Together with noise, these delays can strongly influence both stationary and dynamical properties. They underlie phenomena such as phantom traffic jams~\cite{bando1998analysis,TrafficJamsReview}, bacterial chemotaxis~\cite{deGennes2004}, and the emergence of optimal propulsion speeds in feedback-controlled active Brownian microswimmers~\cite{muinos2021reinforcement}.
    }
    \label{fig:Advertisement}
\end{figure}

Interacting with the world generally requires sensing its state and adjusting actions accordingly (see Fig.~\ref{fig:Advertisement}). This feedback-based regulation is central to control theory~\cite{wiener1948cybernetics, Astrom2010feedback}, which originated in engineering but now plays an essential role across many areas of science. In biology, diverse feedback mechanisms govern homeostasis, regulation, reflexes, and complex action–reaction loops~\cite{cannon1932wisdom, Doyle2011, kandel2000principles}. Similarly, modern physics increasingly relies on explicit measurement and control strategies as experimental techniques have advanced~\cite{bechhoefer2021control}.

In particular, the development of micromanipulation methods—such as optical tweezers~\cite{Ashkin86}, atomic force microscopy~\cite{Binnig1986}, and trapping techniques for cold atoms and ions~\cite{CohenTannoudji1998}—made it possible to implement real-time feedback directly at the microscale. These tools enabled both feedback coupling between microscopic degrees of freedom~\cite{Khadka2018} and the construction of “virtual” or effective potentials generated entirely by feedback algorithms~\cite{bechhoefer2021control}. Such controlled microscale systems have become important testbeds for nonequilibrium statistical physics, including experimental explorations of stochastic thermodynamics, generalized Landauer bounds, and fluctuation theorems~\cite{Collin2005,Berut2012,seifert2025stochastic}.

More recently, feedback has become central to active-matter research~\cite{teVrugt2025, FeedbackAMReview}, where self-driven particles continuously consume energy to regulate their motion, giving rise to structures and behaviors absent in equilibrium systems. This is largely due to the natural feedback interactions observed in biologically active matter—such as in bacteria, animals, or humans—as well as ongoing efforts to replicate and study these systems in the laboratory.

Control of artificial active-matter systems involves two features uncommon in standard engineering problems. First, their many-body nature requires the simultaneous control of large numbers of identical units that may also interact via feedback, whereas typical engineering systems (e.g., vehicle speed control) are low-dimensional. Second, especially at the microscale, noise plays a dominant role. Although noise is present in all systems, conventional engineering seeks to suppress it through isolation or damping—strategies that are largely impractical for bacteria or synthetic microswimmers. As a result, active-matter control often aims not to eliminate noise but to harness it as a functional resource~\cite{Goychuk2016}.

As illustrated in Fig.~\ref{fig:Advertisement}, feedback control is inherently coupled with perception--action delays. In control theory, such delays have been used to stabilize periodic orbits in chaotic systems~\cite{ControlofChaos}, but they are more often regarded as a spurious source of oscillations, instabilities, and generally poor control performance. This is evidenced by their role as a main cause of so-called phantom traffic jams~\cite{bando1998analysis,TrafficJamsReview} and, more broadly, of car accidents -- indeed, the strict rule ``don't drink and drive'' exists because alcohol increases a driver's reaction time.

The simultaneous presence of noise and feedback delay in active matter systems remains relatively understudied~\cite{atay2010complex}. From a mathematical standpoint, these systems are governed by stochastic delay differential equations, which are notoriously difficult to treat analytically beyond approximate approaches such as linearization or short- and long-delay expansions~\cite{loos2021stochastic}. Recent experimental and theoretical work nevertheless reveals a recurring theme: the interplay of noise and delay gives rise to novel phenomenology, affecting both stationary properties and dynamical phases. Moreover, the delay itself can serve as the sole control parameter, enabling switches between these patterns. To date, such behavior has been demonstrated only in a limited set of model systems reviewed below. A general theoretical framework has yet to be established, with the main obstacle being the analytical intractability of stochastic delay differential equations.

\section{Static patterns in systems with spatially varying activity}

Consider an active Brownian particle, either a bacterium or a microswimmer, that swims in a complex environment, making its speed $v(\textbf{x})$ a function of its position $\textbf{x}$. Such speed modulation has been shown experimentally to enhance gradient climbing and population accumulation in chemokinetic bacteria~\cite{Kwangmin2016,Jakuszeit2021}.

For simplicity, let us now focus on two-dimensional setups, where the particle dynamics is described by the equation
\begin{eqnarray}
 \dot{\textbf{x}}(t) &=& v(\textbf{x}(t-\tau)) \textbf{n}(t) + \sqrt{2D}\boldsymbol{\xi}(t),\label{eq:xAL}\\
 \dot{\theta}(t) &=& \sqrt{2D_r}\xi_r(t),\label{eq:tht}
\end{eqnarray}
 where $D$ and $D_r$ are translational and rotational diffusivities corresponding to independent normalized white noises $\boldsymbol{\xi}(t)$ and $\xi_r(t)$, $\textbf{n}(t) = [\cos(\theta), \sin(\theta)]$ denotes the swimming direction of the particle, and $\tau$ is perception-reaction delay with which the particle adjust its speed to that imposed by the landscape.

The classical result by Snitzer~\cite{Schnitzer1993} shows that for vanishing $D$ and $\tau$ the stationary density of an ideal gas of such particles is inversely proportional to their local speed, $\rho(\textbf{x}) \sim 1/v(\textbf{x})$.

\begin{figure}
    \centering
    \includegraphics[width=0.45\textwidth]{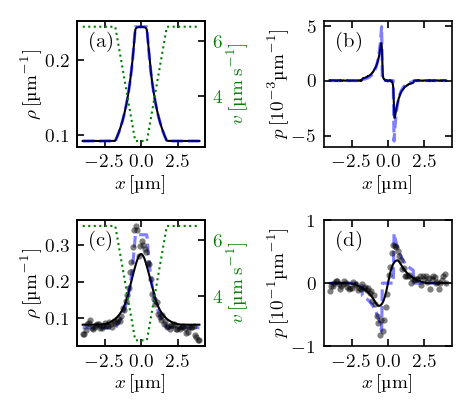}
    \caption{Density, $\rho$, (a,c) and polarization, $p$, (b,d), corresponding to the one-dimensional velocity profile $v$ shown in panels (a,c), for vanishing delay (a,b) and a delay of $\SI{184}{\milli\second}$ (c,d). Experimental results obtained with thermophoretic microswimmers~\cite {Franzl2021} are shown as symbols, the blue dashed line denotes theoretical predictions in Eqs.~\eqref{eq:rho} and \eqref{eq:p}, and the solid black line corresponds to simulations. Parameters: $D_\theta = 2.9\,\Drunit$, $D = 0.04\,\Dunit$. Data from Ref.~\cite{holubec2025delayedactiveswimmervelocity}.
    }
    \label{fig:polarization_inversion}
\end{figure}

In the presence of a nonzero translational diffusivity and short delays ($D_r\tau \ll 1$ and $D\tau \ll 1$), this result has been generalized to~\cite{holubec2025delayedactiveswimmervelocity}
\begin{equation}
\rho = \rho_0 \left(\frac{1}{v^2 + 2 D D_r}\right)^{\frac{1+D_r\tau}{2}}.
\label{eq:rho}
\end{equation}
Also it was shown that the average polarization of the system is given by
\begin{equation}
p = \frac{v^2\tau^2 - 2D\tau}{v^2+2 D D_r}\frac{\rho}{2\tau} \nabla v.
\label{eq:p}
\end{equation}
For $D=0$, the result for the density has been verified in experiments with light-controlled robots~\cite{mijalkov2016engineering,leyman2018tuning}. In the Brownian regime ($D>0$), both results were verified by Brownian-dynamics simulations and by experiments on single thermophoretic microswimmers~\cite{holubec2025delayedactiveswimmervelocity}, as shown in Fig.~\ref{fig:polarization_inversion}.

The main observations are that delay enhances localization in low-velocity regions and that it controls both the magnitude and the sign of polarization. Importantly, the phenomenology survives into the dilute many-body regime. Hence, in addition to controlling density and polarization patterns by tailoring activity fields—which can be experimentally challenging—one can therefore also regulate these patterns through the temporal programming of the time delay. Such regulation can find applications in programmable active metamaterials and reconfigurable photonic media, where the polarization and density could be associated with electric or magnetic polarization, or with optical properties of the active material.

The above discussion applies to situations where transport is suppressed either by the design of the experimental setup or by symmetry. When these constraints are relaxed, time delay can additionally be used to induce and control transport in such systems~\cite{rein2025}.

\section{Delay-induced activity}

\begin{figure}
    \centering
    \includegraphics[width=0.45\textwidth]{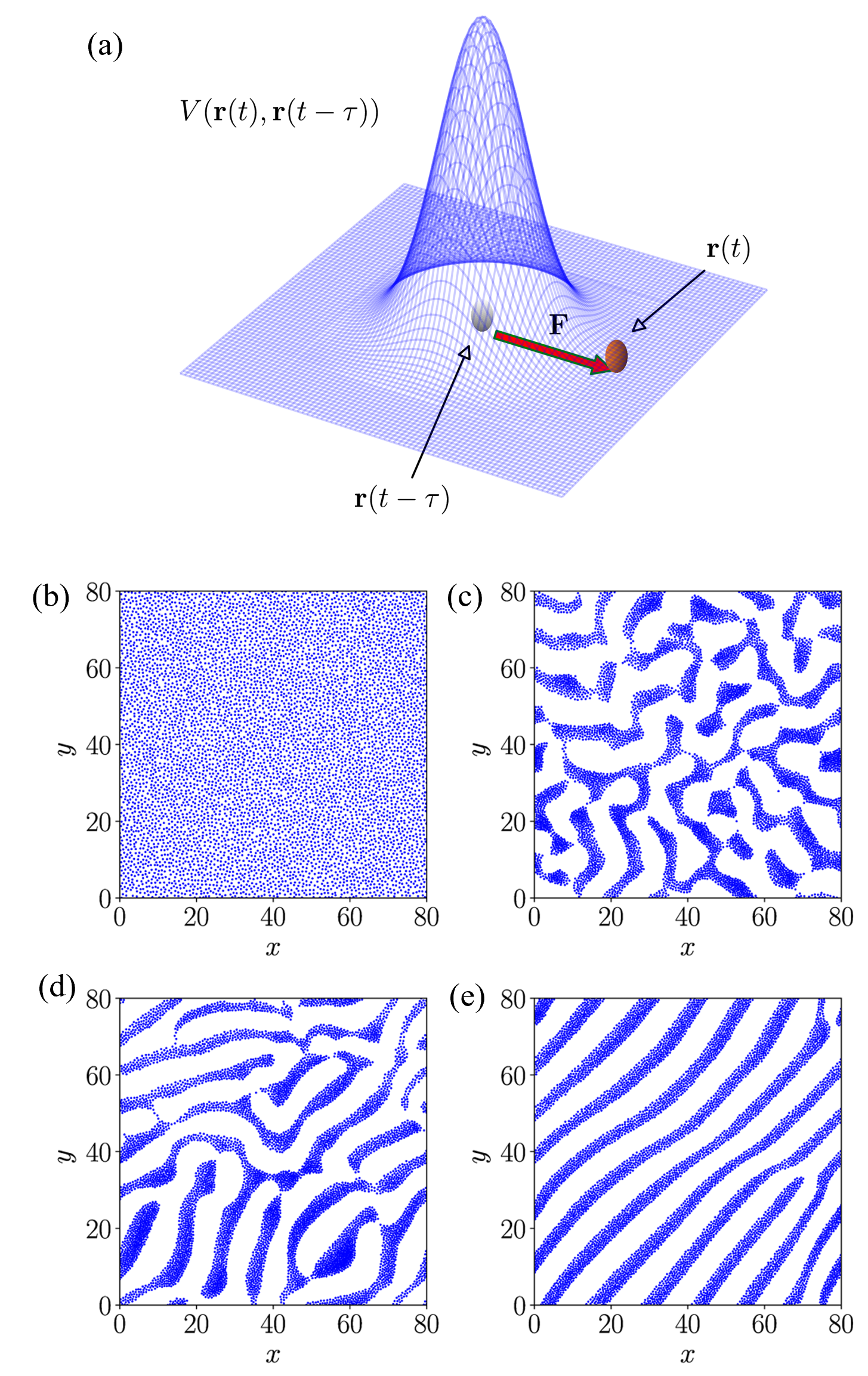}
    \caption{(a) A particle repelled from its own past position and subject to noise undergoes a symmetry-breaking transition, resulting in an active state. Figure reprinted with permission from R. A. Kopp and S. H. L. Klapp, Phys. Rev. E, 107, 024611 (2023). Copyright (2023) by the American Physical Society~\cite{Klapp2023}.
    (b–e) Formation of bands in a system of sterically interacting, passive Brownian particles experiencing strong delayed repulsion from their past positions. Panels (b–e) show system snapshots at times $t = 0, 10, 50$, and $100$, respectively. Simulation results reprinted with permission from S. Tarama, S. U. Egelhaaf, and H. Löwen, Phys. Rev. E, 100, 022609 (2019). Copyright (2019) by the American Physical Society~\cite{Lowen2019}.
    }
    \label{fig:BandsLowen}
\end{figure}

When a passive Brownian particle is controlled by a feedback mechanism that repels it from its own past position, the dynamical equation~\eqref{eq:xAL} changes to
\begin{equation}
 \dot{\mathbf{x}}(t) = F\!\left[\mathbf{x}(t)-\mathbf{x}(t-\tau)\right] + \sqrt{2D}\,\boldsymbol{\xi}(t),
 \label{eq:xSD}
\end{equation}
with \(\operatorname{sgn}[F(x)] = \operatorname{sgn}(x)
\). In general, such a particle undergoes a spontaneous symmetry-breaking transition into an active state, with a self-propulsion speed \(v\) given by the solutions of \(F(v\tau)-v=0\)~\cite{Klapp2023}, as shown in Fig.~\ref{fig:BandsLowen}a.

This type of system has so far been investigated mostly theoretically~\cite{BellDavies2023}, in various many-body generalizations involving finite-size particles. Interestingly, when a particle is repelled not only from its own past position but also from the past positions of its neighbors within an interaction radius, the resulting phenomenology resembles that of the Vicsek model, exhibiting aligned, banded, and chaotic phases~\cite{Kopp2023EPL}. A similar phenomenology persists even when the self-repulsion term is absent~\cite{Lowen2019}, as shown in Fig.~\ref{fig:BandsLowen}b. Most surprisingly, activity can be sustained even in many-body systems with retarded attractive interactions, which do not induce activity at the single-particle level~\cite{Tarama2025}.

\section{Delay-induced chirality}

Consider an active Brownian particle swimming at a constant speed $ v_0 $ toward a fixed target. If the feedback loop controlling the swimming direction involves a time delay $ \tau $, the dynamical equation~\eqref{eq:xAL} is modified to
\begin{equation}
 \dot{\mathbf{x}}(t) = - v_0 \hat{\mathbf{x}}(t-\tau) + \sqrt{2D}\,\boldsymbol{\xi}(t),
 \label{eq:xRot}
\end{equation}
where $ \hat{\mathbf{x}} = \mathbf{x}/|\mathbf{x}| $. The feedback force can be rewritten as $v_0\big[\hat{\mathbf{x}}(t)-\hat{\mathbf{x}}(t-\tau)\big] - v_0 \hat{\mathbf{x}}(t)$, which resembles a combination of a ``repulsion'' from the particle's own past trajectory and an effective V-shaped potential $ v_0 |\mathbf{x}(t)| $. The potential force $- v_0 \hat{\mathbf{x}}(t) $ acts purely in the radial direction. Consequently, upon projection onto the tangential direction---i.e., for the angular velocity---one recovers an equation of the form~\eqref{eq:xSD} with an effective repulsive force $F$.

\begin{figure}
    \centering
    \includegraphics[width=0.45\textwidth,trim=20 72 580 70,clip]{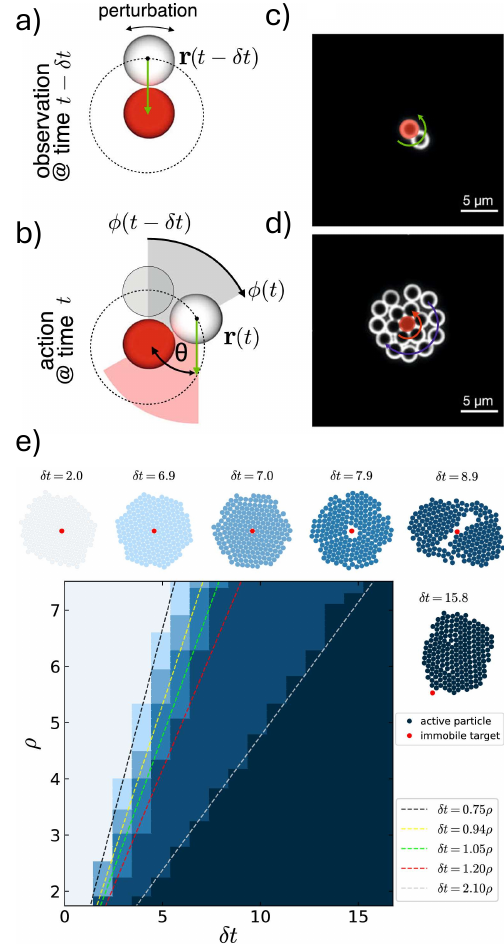}
    \caption{Retarded attraction induces chirality.
(a) A particle locates the target at time $t-\delta t$, and
(b) begins swimming toward it with fixed speed $v_0$ at time $t$.
(c) When the product $\delta t v_0$ is sufficiently large, the particle transiently rotates around the target.
(d) Multiple particles attracted to the same target synchronize their rotation due to steric and hydrodynamic interactions. Figures a--d reprinted from X.~Wang et al., Spontaneous vortex formation by microswimmers with retarded attractions. Nature Communications 14, 56 (2023). Copyright (2023) by The Authors. Licensed under \href{https://creativecommons.org/licenses/by/4.0/}{CC BY 4.0}~\cite{Wang2023}.
(e) For large numbers of particles interacting only via steric forces, increasing the delay at fixed density $\rho$ generates a rich sequence of dynamical phases.
The system transitions from a static crystallite to a series of rotating states: crystallite, crystallite with shear bands, circle, yin–yang, and blob.
Figure reprinted
from P.-C. Chen et al., Active particles with delayed attractions form quaking crystallites. EPL 142, 67003 (2023). Copyright (2023) by The Authors. Licensed under \href{https://creativecommons.org/licenses/by/4.0/}{CC BY 4.0}~\cite{Chen2023}.
}
    \label{fig:ShearBands}
\end{figure}

The feedback-induced attraction thus leads to spontaneous chiral symmetry breaking and the emergence of transiently rotating states~\cite{Wang2023}. As illustrated in Fig.~\ref{fig:ShearBands}a--b, this behavior can be interpreted in terms of delay-induced aiming errors, providing a microscopic basis for corresponding phenomenological chiral symmetry-breaking theories~\cite{Bauerle2020}.

The effective potential governing the angular velocity is controlled by the product \(v_0\tau\) and undergoes a pitchfork bifurcation into two symmetric rotating states at \(\tau=1\). Remarkably, the main dynamical features persist in many-body systems of such particles attracted to a common center and interacting either via excluded-volume interactions in simulations~\cite{Chen2023} or via hydrodynamic--thermophoretic interactions in experiments~\cite{Wang2023} (see Fig~\ref{fig:ShearBands}c-e).

However, the many-body dynamics is richer: interactions can synchronize the motion of individual particles, and hydrodynamic--thermophoretic couplings can even generate counter-rotating shells around the target. The resulting phenomenology is illustrated in Fig.~\ref{fig:ShearBands}d--e.

Even more complex behavior arises when many such rotators---each attracted with a delay to its own center, whose dynamics is itself coupled to the particle position---interact via hydrodynamic couplings. This system provides a physical realization of an ensemble of swarmalators, i.e., agents that combine swarming with synchronization~\cite{Heuthe2025}.


\section{Car traffic}

The preceding sections focused on delay effects at the single-particle level and in small clusters. Historically, however, delay-induced collective phenomena were first analyzed in car traffic~\cite{TrafficJamsReview}. There, above a critical density---when the typical headway becomes comparable to the product of the average vehicle speed and the reaction delay---the uniform flow destabilizes into coexisting free-flow and jammed phases~\cite{bando1998analysis}. Because traffic is essentially one-dimensional, it is less suited for studying the richer structure formation observed in bird flocks or insect swarms. This motivated the development of higher-dimensional flocking models reviewed next.

\section{Flocking models}

The effects of time delay in flocking and swarming have been studied mainly within three classes of models. The first is the original Vicsek model with nonlinear alignment interactions, which can be viewed as a moving-spin variant of the Heisenberg model in which spins self-propel at constant speed~\cite{Vicsek1995}. The second is the Cucker--Smale model, introduced to study mathematical conditions for the emergence of consensus in many-body systems~\cite{cucker2007emergent,ShuTadmor2021}. The third class comprises inertial flocking models with linear interparticle couplings, originally introduced by Mikhailov and Zanette~\cite{mikhailov1999noise} and later extended to include explicit time delays by Forgoston and Schwartz~\cite{forgoston2008delay}.

\begin{figure}[t]
\centering
\includegraphics[width=1.0\linewidth]{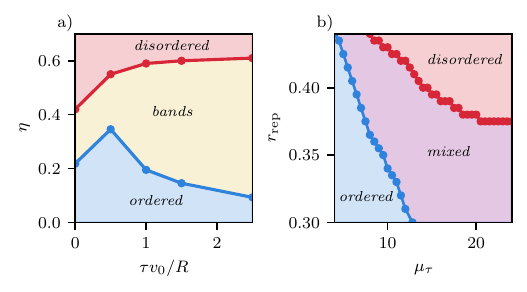}
\caption{
(a) Phase diagram of the delayed Vicsek model as a function of the noise strength $\eta$ and the time delay $\tau$, rescaled by the ratio of agent speed $v_0$ to interaction radius $R$, based on data from R.~Horton and V.~Holubec, Order-disorder transition and phase separation in delay Vicsek
model. New.~J.~Phys.~27, 094402 (2025). Copyright (2025) by The Authors. Licensed under \href{https://creativecommons.org/licenses/by/4.0/}{CC BY 4.0}~\cite{horton2025}.
(b) Phase diagram of a modified delayed Vicsek model as a function of the mean delay $\mu_\tau$ and the strength of repulsive interactions $r_{\mathrm{rep}}$, using data digitized from Fig.~6 of Physica A 634, F. Pakpour and T. Vicsek, Delay-induced phase transitions in active matter, 129453, Copyright (2024), with permission from Elsevier~\cite{pakpour2023delays}.
}
\label{fig:VMstripes}
\end{figure}

A robust trend across flocking models is that short delays tend to promote order, whereas long delays promote disorder~\cite{erban2016cucker,PiwowarczykVicsek2019,HolubecScaling2021,horton2025}. Intuitively, short delays allow agents to effectively average out fluctuations, stabilizing collective motion, whereas long delays impede their ability to track emerging collective trends and favor disordered dynamics. More generally, delay can serve as a control parameter in the phase diagrams of these systems, enabling transitions between distinct dynamical states~\cite{forgoston2008delay,pakpour2023delays,horton2025,ChenZheng2024}. Figure~\ref{fig:VMstripes} illustrates these effects for two variants of the delayed Vicsek model: the original model with a fixed delay~\cite{horton2025} and a variant with stochastic delay and repulsive interactions~\cite{pakpour2023delays}.

In the model introduced by Forgoston and Schwartz~\cite{forgoston2008delay}, the system can display a strongly polarized translating state, a ring state, and---for nonzero delay---a rotating state. Here, too, the time delay serves as a control parameter that drives transitions between these collective states. Because the interactions are global and linear, the dynamics admits a low-dimensional reduction, enabling an analytic determination of the phase diagram. By contrast, a comparable analytic characterization of delay-extended Vicsek-type models---with nonlinear local interactions and fluctuations strongly coupled to density---remains beyond current theoretical methods.

In the original delayed Vicsek model, introducing a time delay modifies how information propagates, changing the dispersion relation from diffusive---where the distance \(d\) traveled over time \(t\) scales as \(d\propto \sqrt{t}\)---to ballistic, \(d\propto t\)~\cite{geiss2022delay}. Time delay also reshapes temporal correlations, turning an overdamped (purely exponential) decay into an underdamped form with an initially flat correlation plateau~\cite{HolubecScaling2021}.

Moreover, the dynamical scaling exponent relating spatial and temporal correlations is about \(2\) at low speeds and \(1.5\) at high speeds in the standard Vicsek model~\cite{Cavagna2023}, whereas it is close to \(1.1\) in the delayed Vicsek model~\cite{HolubecScaling2021}. Taken together, these results suggest that adding a time delay brings the Vicsek model closer to the behavior observed in natural swarms and flocks~\cite{cavagna2018physics}, in a way reminiscent of Vicsek-type generalizations that incorporate orientational inertia~\cite{Cavagna2023,Caprini2021inertial}. However, the two mechanisms are physically distinct: inertia couples to translational degrees of freedom and vanishes in overdamped systems, whereas delay acts on the information flow and persists at any damping. In real swarms and flocks, both effects likely coexist, but in overdamped systems such as bacterial colonies or ensembles of artificial microswimmers, only delay-induced effects remain.

Additionally, the delayed Vicsek model has been applied to study the collective motion of marching locusts in a circular arena~\cite{MarchingLocustsScience2006}. These studies demonstrate that, in certain situations, it is crucial to distinguish between response and transmission delays—a distinction that is often not made in modeling collective motion. In particular, the transition rates between the two ordered states, corresponding to clockwise and counterclockwise motion, depend sensitively on the type of delay: while increasing the response delay always suppresses switching, the effect of transmission delay can be either stabilizing or destabilizing, depending on the response delay~\cite{RatesDelayLocusts2022}.

This observation is further supported by recent mathematical results for Cucker--Smale-type models. While response (processing) delays allow for a well-defined mean-field limit, transmission delays generally do not~\cite{haskovec2024graph}. The reason is that transmission delays introduce self-delay terms that require following individual characteristics of the underlying transport equation, thereby invalidating the usual indistinguishability-of-particles assumption.

Naturally, when moving beyond idealized models toward practical engineering applications—such as the control of swarms of flying drones—feedback delays must be accounted for alongside the full complexity of the underlying physical constraints~\cite{Vasarhelyi2018}.

\section{Outlook}

In classical control theory, feedback delays are typically viewed as nuisances that degrade performance. In contrast, recent work suggests that in noisy active-matter systems, delays can generate new functionalities and serve as a readily tunable control parameter for switching between behaviors and dynamical phases. So far, these conclusions are supported primarily by theoretical studies of low-dimensional models, experiments with up to a few hundred particles, and large-scale simulations of a limited range of systems.

Several delay-induced phenomena, such as delay-induced activity (Fig.~\ref{fig:BandsLowen}) and chirality (Fig.~\ref{fig:ShearBands}), as well as short-delay-enhanced order and long-delay-induced disorder (Fig.~\ref{fig:VMstripes}) appear robust and likely extend beyond the specific models studied so far. By contrast, the generality of other effects remains unclear. In particular, the influence of delay on density, polarization, and transport in active Brownian particle systems, and its impact on (finite-size) scaling exponents and information transfer in flocking models, calls for more systematic exploration.

The delay dependence of static observables including density profiles, polarization, and potentially higher-order moments in many-body settings opens the door to dynamically reconfigurable materials whose magnetization or optical properties can be controlled \emph{in situ} by tuning the delay time. Since time delays are among the easiest parameters to adjust in feedback-control experiments, we expect substantial research activity in this direction.

A closely related opportunity is to use delay to control transport. Our results indicate that delay-induced currents in spatially periodic activity landscapes can reach rectification efficiencies (ratio of directed to total speed) of up to \(\sim 50\%\) already at the single-particle level~\cite{rein2025}. We further find that this effect is robust to the specific form of the memory kernel and persists for exponential delays, which are ubiquitous in biology (e.g., bacterial chemotaxis~\cite{deGennes2004}). The corresponding effects in many-body systems remain largely unexplored.

The main theoretical bottleneck in analyzing higher-dimensional systems is the notorious analytical intractability of the underlying stochastic delay differential equations. These generally permit analysis only via long-delay approximations~\cite{Lowen2019}, short-delay expansions~\cite{holubec2025delayedactiveswimmervelocity}, or suitable linearizations~\cite{HolubecScaling2021}. The root of this difficulty lies in their infinite-dimensional nature. Consequently, for nonlinear systems the probability-density dynamics is not closed, precluding a closed Fokker--Planck—and thus hydrodynamic—description beyond such approximations. Since hydrodynamic theories are central to symmetry-based approaches to collective behavior and scaling, a key challenge is to develop a controlled approximation scheme that yields a tractable yet qualitatively accurate hydrodynamic theory for active systems with delay~\cite{teVrugtJerky2021}.

Another promising direction concerns \emph{intrinsic} physico-chemical response delays in nonstationary environments. For instance, when a catalytically active particle traverses regions with spatially varying fuel concentration, its activity may adjust only after a finite response time set by reaction kinetics and, depending on the propulsion mechanism, by thermophoretic or diffusiophoretic relaxation~\cite{Topfer2025}. Such intrinsic delays can generate autonomous feedback between motion and the surrounding chemical field, enabling nontrivial functionalities including motion against chemical gradients, which is particularly appealing for applications such as targeted \emph{in vivo} drug delivery.

More broadly, response delays can be viewed as a dynamical ingredient of \emph{morphological intelligence}. While the term is often associated with static structure, the examples reviewed in this Perspective motivate extending it to \emph{temporal morphology}, in which intrinsic timescales, such as delay times, enable functional responses. In this sense, a particle migrating up a gradient due to reaction-kinetic delays is not explicitly ``sensing'' or ``computing'' the gradient; it is exploiting its own response dynamics.  This viewpoint unifies several seemingly disparate phenomena.

\emph{Reservoir computing.}
Delay systems are a canonical substrate for physical reservoir
computing, where high-dimensional nonlinear dynamics maps inputs to
outputs without weight training in the
reservoir~\cite{Appeltant2011}.  In our context, the delayed
rotational dynamics of microswimmers attracted to a common target (Fig.~\ref{fig:ShearBands})
already provides the requisite nonlinear mixing and fading memory~\cite{Wang2024,BechingerRC}.
Morphological intelligence here means that computation is performed by
the physical dynamics itself, not by an external algorithm.

\emph{Bacterial chemotaxis.}
The exponential memory kernel underlying sensory adaptation in
\textit{E.~coli}~\cite{deGennes2004} implements temporal comparisons
over a characteristic timescale of $\sim 1$\,s---long enough to average
over tumbles, short enough to track gradients.  This biochemical delay
is precisely what enables gradient climbing in noisy environments where
instantaneous measurements are useless.  It is morphological
intelligence realized through the kinetics of the chemotactic signaling
cascade.

\emph{Embodied intelligence.}
Recent experiments have demonstrated that synthetic microswimmers can
adapt to hidden hydrodynamic perturbations through physical embodiment
alone, without any explicit flow sensing~\cite{Paul2026}.  Using
reinforcement learning to control self-thermophoretic particles, Paul
\textit{et~al.} showed that the correlations between actions, physical
dynamics, and outcomes encode sufficient environmental information for
successful navigation---even when the perturbation is entirely absent
from the agent's state representation.  This provides direct
experimental evidence that body--environment interactions can replace
sensors, and suggests that intrinsic delay is one specific mechanism
through which such embodied computation operates.

\emph{Stigmergy.}
Delay-mediated self-organization---where agents respond to their own
past states~\cite{Lowen2019,Klapp2023} or to traces left by
others~\cite{StigmergyReview,Wang2023}--represents a form of
environment-mediated coordination.  The delay effectively outsources
memory to the interaction history, enabling collective behavior without
direct communication--a form of morphological intelligence
embedded in the environment rather than in the agents themselves.

Together, these examples reveal that time delay is a fundamental and
underexplored dimension of morphological intelligence in active matter.
The systems reviewed here---from single delayed microswimmers to
delay-coupled flocks---suggest that the interplay of noise, delay, and
physical dynamics constitutes a general-purpose information-processing
resource, one that evolution has exploited extensively and that
artificial active-matter design is only beginning to harness.


\emph{Data availability statement:} No new data were created
or analysed in this study.

\printbibliography
\end{document}